\documentclass[copyright]{eptcs}
\providecommand{\event}{SYNT 2015} 
\usepackage{breakurl}   

\usepackage{versions}

\usepackage{amssymb}
\usepackage{amsmath}
\usepackage{txfonts}
\usepackage{amssymb}
\usepackage{enumerate}
\usepackage{amsfonts}
\usepackage{times}
\usepackage{mathrsfs}
\usepackage{amscd}
\usepackage{stmaryrd}
\usepackage{graphicx}
\usepackage{makeidx}
\usepackage{multirow}
 \usepackage{fancyvrb}

\newcommand{\commentout}[1]{}

\newcommand{\Points}{\mathit{Points}}
\newcommand{\act}[1]{\mathbf{#1}}

\newcommand{\CTL}{\mbox{CTL}}
\newcommand{\CTLK}{\mbox{CTLK}}

\newcommand{\I}{\mathcal{I}}
\newcommand{\R}{\mathcal{R}}

\newcommand{\nat}{\mathbf{N}}

\newcommand{\rimp}{\Rightarrow}
\newcommand{\dimp}{\Leftrightarrow}

\newcommand{\Ags}{Ags}

\newcommand{\trans}{\longrightarrow}
\newcommand{\ptrans}[1]{\stackrel{#1}{\longrightarrow}}

\newcommand{\powerset}[1]{{\cal P}(#1)}
\newcommand{\until}{\!U\!}

\newcommand{\Prop}{\mathit{Prop}}

\newtheorem{lemma}{Lemma}

\newtheorem{theorem}{Theorem} 
\newtheorem{example}{Example} 
 
\newtheorem{proposition}{Proposition}

\newcommand{\qed}{$\boxempty$}

\newcommand{\be}{\begin{enumerate}} 
\newcommand{\ee}{\end{enumerate}}

\newcommand{\Prot}{\mathit{Prot}}

\newcommand{\KProt}{\mathtt{P}}

\newcommand{\Var}{V}
\newcommand{\Vars}{\mathit{Vars}}

\newcommand{\Acts}{\mathit{Acts}}

\newcommand{\pdo}{\mathbf{do}} 
\newcommand{\pdor}{\mathbf{od}}

\newcommand{\Spec}{\Phi} 
\newcommand{\Env}{E} 

\newcommand{\EPS}{{\cal S}}
\newcommand{\enabled}{\mathit{en}}

\usepackage{hyperref}
\begin{document}

\title{The complexity of approximations for epistemic synthesis (extended abstract)}

\author{Xiaowei Huang
\institute{UNSW Australia}
\email{xiaoweih@cse.unsw.edu.au}
 \and 
Ron van der Meyden
\institute{UNSW Australia}
\email{meyden@cse.unsw.edu.au}
}
\def\titlerunning{The complexity of approximations for epistemic synthesis (extended abstract)}
\def\authorrunning{X. Huang \& R. van der Meyden}

\maketitle

\newcommand{\reach}{rch}
\newcommand{\Reach}{Rch}
\newcommand{\sat}{sat}
\newcommand{\Sat}{Sat}

\newcommand{\states}[1]{{\cal P}(#1)}

\begin{abstract}
Epistemic protocol specifications allow programs, for 
settings in which multiple agents act with incomplete information, to be 
described in terms of how actions are related to what the agents know. 
They  are a variant of the knowledge-based programs of  Fagin et al [Distributed Computing, 1997],
motivated by the complexity of synthesizing implementations in that framework. 
The paper proposes an approach to the synthesis of implementations of 
epistemic protocol specifications, that reduces the problem of finding an implementation to 
a sequence of model checking problems in approximations of the ultimate system being synthesized. 
A number of ways to construct such approximations is considered, and these are 
studied for the complexity of the associated model checking problems. 
The outcome of the study is the identification of the best approximations with the 
property of being PTIME implementable. 

\end{abstract}

\section{Introduction}
\label{sec:intro} 

\emph{Knowledge-based programs} \cite{FHMV1997} are an abstract specification format for
concurrent systems, in which the actions of an agent are conditional on formulas of the logic of knowledge \cite{FHMVbook}. 
This format allows the agent to be described in terms of  
what it must know in order to perform its actions, independently 
of how that knowledge is attained or concretely represented by the agent.  This 
leads to implementations that are optimal in their use of the knowledge implicitly available in an agent's local state. 
The approach has been applied to  problems including reliable message transmission \cite{HZ92}, 
atomic commitment \cite{Had87}, fault-tolerant agreement \cite{DM90}, robot motion planning \cite{BrafmanLMS1997} and cache coherency \cite{BaukusMeyden}. 

The process of going from an abstract knowledge-based program to a concrete implementation is non-trivial, since 
it requires reasoning about all the ways that knowledge can be obtained, 
which can be quite subtle. Adding to the complexity, there is a circularity in that knowledge 
determines actions, which in turn affect the knowledge that an agent has. 
It is therefore highly desirable to be able to automate the process of
implementation. 
Unfortunately, this is known to be an inherently complex problem: even 
deciding whether an implementation exists is intractable \cite{FHMV1997}. 

Sound local proposition epistemic specifications \cite{EMM98} 
are a generalization of knowledge-based programs proposed in part 
due to these complexity problems. These specifications require only 
\emph{sufficient} conditions for knowledge, where knowledge-based programs
require \emph{necessary and sufficient conditions}. 
 By allowing a larger space of potential implementations, 
this variant ensures that there always exists an implementation. However, some of
these implementations are so trivial as to be uninteresting. 
In practice, one wants implementations in which agents make good use of their
knowledge, so that the conditions under which they act closely approximate
the necessary and sufficient conditions for knowledge. 
To date, 
a systematic approach to the identification of \emph{good} implementations, and 
of automating the construction of such good implementations, has not been identified. 
This is the problem we address in the present paper.  Ultimately, we seek an automated approach that is 
implementable in a way that scales to handling realistic examples. In this paper, we 
use a CTL basis for specifications, and use PTIME  
complexity of an associated model checking problem in an 
explicit state representation as a proxy for 
practical  
implementability. 

The contributions of the paper are two-fold: first, we present a general approach
to the identification of good implementations, that extends the 
notion of sound local proposition epistemic specification by ordering the 
knowledge conditions to be synthesized, and then defining a 
way to construct implementations using a sequence of approximations
to the final synthesized system, in which implementation choices for earlier 
knowledge conditions are fed back to improve  the quality of approximation
used to compute later implementation choices.  
This gives an intuitive approach to the construction of implementations, which 
we show by example to address some unintuitive aspects of the original
knowledge-based program semantics. The approach is parametric in a choice of approximation scheme. 

Second, we consider a range of possibilities for the approximation scheme
to be used in the above ordered semantics, and 
evaluate the complexity of the synthesis computations associated with 
each approximation. The analysis leads to the identification of two
orthogonal approximations that are optimal in their closeness to 
a knowledge-based program semantics, while remaining PTIME 
computable. This identifies the best prospects for future work on 
synthesis implementations.

The paper is structured as follows. 
Section~\ref{sec:mckt} recalls basic definitions of temporal epistemic logic. 
Section~\ref{sec:eps} defines epistemic protocol specifications.  
In Section~\ref{sec:ordsem} we define the ordered semantics approximation approach 
for identification of good implementations. Section~\ref{sec:approx} 
defines a range of possible approximation schemes, which 
are then analyzed for complexity in Section~\ref{sec:complex}.  
We discuss related work in  Section~\ref{sec:related}
 and conclude with a discussion of future work in Section~\ref{sec:concl}.

\section{A Semantic model for Knowledge and Time}
\label{sec:mckt}

In this section we lay out a general logical framework for agent knowledge, 
and describe how knowledge arises for agents that execute a concrete
protocol in the context of some environment. 

 Let $\Prop$ be a finite set of atomic propositions and $\Ags$ be a finite set of agents. The language  CTL$^*$K$(\Prop,\Ags)$ 
 has the syntax: 
$$\phi::= ~p~|~\neg \phi~|~\phi_1\lor \phi_2~|~X\phi~|~(\phi_1\until\phi_2)~|~A\phi~|~K_i\phi$$
where $p\in \Prop$ and $i\in\Ags$.  
This is CTL$^*$ plus the construct $K_i \phi$, which says that agent $i$ knows that $\phi$ holds. 
We freely use standard  operators that are definable in terms
of the above, specifically $F \phi  = \mathbf{true} \,\until \phi$,  $G \phi = \neg F \neg \phi$, 
$\phi_1 R \phi_2 = \neg ((\neg \phi_1)\, \until \,(\neg \phi_2))$, $E\phi = \neg A \neg \phi$. 
Our focus in this paper is on the fragment $\CTLK$, in which 
the branching operators may occur only as $A\phi$ and $E\phi$, 
where $\phi$ is a formula in which the outermost operator is
one of the temporal operators $X, \until, R, F$ or $G$. 
A further subfragment of this language $\CTLK^+$, 
specified by the grammar 
$$\phi::= ~p~|~\neg p~|~\phi_1\lor \phi_2~|~\phi_1 \land \phi_2~|~AX\phi~|~AF\phi~|~ AG\phi~|~ A(\phi_1\until\phi_2)~|~A(\phi_1 R\phi_2)~|~K_i\phi$$
where $p\in \Prop$ and $i\in\Ags$.  
Intuitively, this is the sublanguage in which all occurrences of the operators $A$ and $K_i$ are in 
positive position. 

To give semantics to all these languages it suffices to give semantics to  CTL$^*$K$(\Prop,\Ags)$. 
We do this using a  variant of interpreted systems \cite{FHMVbook}. 
Let $S$ be a set, which we call the set of global states. 
A {\em run} over $S$ is a function $r:\nat \rightarrow S$. 
A {\em point} is a pair $(r,m)$ where $r$ is a run and $m\in \nat$. 
Given a set $\R$ of runs, we define $\Points(\R)$ to be the set of all points of runs 
$r\in \R$. 
An {\em interpreted system} for $n$ agents is a tuple $\I = (\R, \sim, \pi)$, 
where $\R$ is a set of runs over $S$, 
the component $\sim$ is a collection $\{\sim_i\}_{i \in \Ags}$, 
where for each $i \in \Ags$,  $\sim_i$ is an equivalence relation 
on $\Points(\R)$ (called agent $i$'s {\em indistinguishability relation}) 
and $\pi: S\rightarrow \powerset{\Prop}$ is an interpretation function. 
We say that a run $r'$ is {\em equivalent to a run $r$ up to time $m\in \nat$} if 
$r'(k) = r(k)$ for  $0\leq k\leq m$.

We can define a general semantics of  CTL$^*$K$(\Var,\Ags)$  by means of a relation $\I,(r,m)\models \phi$, 
where $\I$ is an intepreted system, $(r,m)$ is a point of $\I$ and $\phi$ is a formula. 
This relation is defined inductively as
follows: 
\begin{itemize} 
\item $\I,(r,m) \models p$ if   $p\in \pi(r(m))$,  for $p \in \Prop$; 
\item 
$\I,(r,m)\models \neg \phi$ if not $\I,(r,m)\models \phi$; 
\item 
$\I,(r,m)\models \phi_1\lor \phi_2$ if $\I,(r,m)\models \phi_1$ or $\I,(r,m)\models \phi_2$; 

\item 
$\I,(r,m)\models A \phi$ if $\I,(r',m)\models \phi$ for all runs $r'\in \R $ equivalent to $r$ up to time $m$;

\item
$\I,(r,m)\models X \phi$ if  $\I,(r,m+1)\models \phi$; 

\item 
$\I,(r,m)\models \phi_1U\phi_2$ if  there exists $m'\geq m$ such that 
$\I,(r,m') \models \phi_2$, and $\I,(r,k)\models \phi_1$ for $m \leq  k < m'$; 

\item 
$\I,(r,m)\models K_{i} \phi$ if  $\I,(r',m' ) \models \phi$ for all  points $(r',m') \sim_i (r,m)$ of $\I$.

\end{itemize} 
For the knowledge operators, this semantics is essentially the same as the
usual interpreted systems semantics. For the temporal operators, 
it corresponds to a semantics for branching time known as the 
{\em bundle semantics} \cite{Burgess,MeydenWong}.  
We write $\I \models \phi$ when $\I,(r,0)\models \phi$ for all runs $r$ of $\I$. 

We are interested in systems in which each of the agents runs a protocol 
in which it  chooses its actions based on local information, in the context of a larger environment. 
An {\em environment} for agents $\Ags$ is a tuple 
$\Env =  \langle S, I, \{\Acts_i\}_{i \in \Ags}, \trans,  \{O_i\}_{i\in \Ags},
\pi\rangle$, where 
\be \item 
$S$ is a finite set of states, 
\item $I$ is a subset of $S$, representing the initial states, 
\item for 
each agent $i$, component $\Acts_i$ is a finite set of actions
that may be performed by agent $i$; we define 
$\Acts = \Pi_{i\in Ags} \Acts_i$ to be the corresponding set of  \emph{joint actions} 
\item $\trans \, \subseteq  S \times \Acts \times S$ is a transition relation, 
labelled by joint actions, 
\item for each $i\in \Ags$, component $O_i$ is a mapping from $S$ to 
some set $O$ of observations, 
\item $\pi: S\rightarrow \powerset{\Prop}$ is an interpretation of some set of atomic propositions $\Prop$. 
\ee
Intuitively, a joint action $\act{a}$ represents a choice of action $\act{a}_i$ for each agent, 
performed simultaneously, and the transition relation resolves this into an effect on the 
state. We assume that $\trans$ is serial in the sense that for all $s\in S$ and 
$\act{a} \in \Acts$ there exists $t\in S$ such that $s\ptrans{\act{a}} t$.
We assume that $\Acts_i$ always contains at least an action $\act{skip}$, 
and that for the joint action $\act{a}$ with $\act{a}_i = \act{skip}$ for all agents $i$, 
we have 
$s\ptrans{\act{a}} t$ iff $s=t$. 
The set $O$ of observations is an arbitrary set: for each agent $i$, we will be 
interested in the equivalence relation 
$s\sim_i t$  if $O_i(s) = O_i(t)$ induced 
by the observation function $O_i$ rather than the actual values of $O_i$. 

A proposition $p$ is \emph{local to agent  $i$} in the enviroment $\Env$ if it depends only on the
agent's observation, in the sense that for all states $s,t$ with $O_i(s) = O_i(t)$,  
we have $p \in \pi(s)$ iff $p \in \pi(t)$. We write $\Prop_i$ for the set of 
propositions local to agent $i$. Intuitively, these are the propositions
whose values the agent can always determine, based just on its observation. 
We similarly say that a boolean formula is local to agent $i$ if it contains 
only propositions that are local to agent $i$. 
We assume that the set of local propositions is \emph{complete} with respect to 
the observations, in that for each observation $o$ there exists a local formula
$\phi$ such that for all states $s$, we have $O_i(s) = o$ iff $\pi(s) \models \phi$. 
(This can be ensured by including a proposition $p_o$ that is  true at just states $s$ with $O_i(s) = o$, 
or by including a proposition $v =c$ for each possible value $c$ of each variable $v$ 
making up agent $i$'s observation.)

A {\em concrete protocol} for agent $i\in \Ags$ in such an environment $\Env$
is a Dijkstra style nondeterministic looping statement $P_i$ of the form 
\begin{equation}\label{eq:prot}  
  \pdo ~~\phi_1 \rightarrow a_1~ []~ \ldots~ []~ \phi_k \rightarrow a_k ~~ \pdor 
  \end{equation} 
where the $a_j$ are actions in $\Acts_i$ and the $\phi_j$ are 
boolean formulas local to agent $i$. 
Intuitively, this is a nonterminating program that is executed by the agent
repeatedly checking which of the  guards $\phi_j$ holds, and then nondeterministically 
performing one of the corresponding actions  $a_i$. 
If none of the guards holds, then the action $\act{skip}$ is performed. 
That is, implicitly, there is an additional clause $\neg \phi_1 \land \ldots \neg \phi_n \rightarrow \act{skip}$. 
Without loss of generality, we may 
assume that the $a_i$ are distinct. 
(We can always amalgamate two cases $\phi_1 \rightarrow a$ and $\phi_2 \rightarrow  a$ with the same action $a$
into a single case $\phi_1 \lor \phi_2 \rightarrow  a$.) 
We say that action $a_j$ is enabled in protocol $P_i$ at state $s$ if $\phi_j$ holds with respect to the assignment $\pi(s)$, 
and write $\enabled(P_i,s)$ for the set of all actions enabled in protocol $P_i$ at state $s$. 

A \emph{joint protocol $P$} is a collection $\{P_i \}_{i\in \Ags}$ of protocols for the individual agents. 
A joint action $a\in \Acts$ is \emph{enabled by $P$ at a state $s$}  if $a_i \in \enabled(P_i,s)$ for all $i \in \Ags$. 
We write $\enabled(P,s)$ for the set of all joint actions enabled by $P$ at state $s$. 

Given an environment $\Env =  \langle S, I, \{\Acts\}_{i \in \Ags}, \trans,  \{O_i\}_{i\in \Ags},
\pi\rangle$ 
and a joint protocol $P$  for the agents in $\Env$, we may construct an interpreted system $\I(\Env, P) = (\R(\Env, P), \sim, \pi)$ 
over global states $S$ as follows. 
The set of runs $\R(\Env, P)$ consists of all runs $r:\nat \rightarrow S$ such that
$r(0) \in I$ and for all $n \in \nat$ there exists $\act{a} \in \enabled(P, r(n))$ such that $r(n) \ptrans{\act{a}} r(n+1)$. 
The component $\sim = \{\sim_i\}_{i\in \Ags}$ is defined by $(r,m) \sim_i (r',m')$ if $O_i(r(m)) = O_i(r'(m'))$, i.e., two points 
are indistinguishable to agent $i$ if it makes the same observation at the corresponding global states; this is 
known in the literature as the \emph{observational} semantics for knowledge.  
The interpretation $\pi$ in the interpreted system   $\I(\Env, P)$ is identical to that in the 
environment~$\Env$. 

Note that in $\I = \I(\Env, P)$, the satisfaction of formulas of the form $K_i\phi$ 
in fact depends only on the observation $O_i(r(m))$. We therefore
may write  $\I, o \models K_i\phi$ for an observation value $o$ to mean 
$\I, (r,m) \models K_i \phi$ for all points $(r,m)$ of $\I$ with $O_i(r(m)) = o$.

\section{Epistemic Protocol Specifications} 
\label{sec:eps}

 Protocol templates generalize concrete protocols by 
 introducing some variables that may be instantiated with
 local boolean formulas  in order to obtain a concrete protocol. Formally, 
 a {\em  protocol template} for agent $i \in \Ags$ is an 
 expression in the same form as~(\ref{eq:prot}), except that 
 the $\phi_j$ are now boolean expressions, not just over the local atomic propositions $\Prop_i$, but may also contain
 boolean variables from an additional set $X$ of \emph{template variables}. We write $\Vars(\Prot_i)$
 for the set of these additional boolean variables that occur in some $\phi_i$.

An {\em epistemic protocol specification} is a tuple 
$\EPS = \langle \Ags, \Env, \{\KProt_i\}_{i\in \Ags}, \Spec \rangle$,  
consisting of 
a set of agents $\Ags$, 
an environment $\Env$ for $\Ags$, 
a collection of protocol templates $\{\KProt_i\}_{i\in \Ags} $ for environment $\Env$, and
a collection of epistemic logic formulas $\Spec$ over the agents  $\Ags$ and atomic propositions $X \cup \Prop$. 
In this paper, we assume $\Spec \subseteq \CTLK(\Ags,X\cup\Prop)$.
We require that $\Vars(\KProt_i)$ and $\Vars(\KProt_j)$ are disjoint when~$i\neq j$. 

Intuitively, the protocol templates in such a specification lay out the abstract structure of some concrete protocols, 
and the variables in $X$ are ``holes" that need to be filled in order to obtain a concrete protocol.  
The formulas in $\Phi$ state constraints on how the holes may be filled: it is required that these formulas
be valid in the model that results from filling the holes. 
 
To implement an epistemic protocol  specification with respect to the observational semantics, 
we need to replace each template variable $v$ in each agent $i$'s  protocol template 
by an expression over the agent's 
local 
variables,  in such a way that the specification 
formulas are satisfied in the model resulting from executing
the resulting standard program. We now formalize this semantics. 

Let $\theta$ be a substitution mapping each 
template variable $x\in \Vars(\KProt_i)$, for $i \in \Ags$,
to a boolean formula local to agent $i$.
We may apply such a substitution to a protocol template $\KProt_i$ in the form (\ref{eq:prot})
by applying $\theta$ to each of the formulas $\phi_j$, yielding 
$$ 
  \pdo ~~\phi_1\theta \rightarrow a_1~ []~ \ldots~ []~ \phi_k \theta \rightarrow a_k ~~ \pdor 
$$ 
which we write as $\KProt_i\theta$. Since the $\phi_j\theta$ contain only propositions in $\Prop_i$, 
this is a concrete protocol for agent $i$. 
Consequently, we obtain a joint concrete protocol $\KProt \theta = \{\KProt_i\theta\}_{i\in \Ags}$, 
which may be executed in the environment $\Env$, 
generating the system $\I(\Env, \KProt\theta)$. 
The substitution $\theta$ may also be applied to the specification formulas in 
$\Spec$. Each $\phi\in \Spec$ is a formula over variables $X \cup \Prop$, 
so $\phi \theta$ is a formula over variables $\Prop$. 
We write $\Spec\theta$ for $\{\phi \theta ~|~ \phi \in \Spec\}$. 
We say that such a substitution $\theta$ provides an {\em implementation} of the epistemic protocol 
specification 
$\EPS$,  
provided  $ \I(\Env,   \{\KProt_i \theta\}_{i\in \Ags}) \models \Spec\theta$.
The problem we study in this paper is the following: given an environment $\Env$ and an
epistemic protocol specification $\EPS$, synthesize
an implementation $\theta$.

\emph{Knowledge-based programs}  \cite{FHMVbook,FHMV1997} are a special case of epistemic protocol specifications. 
Essentially, knowledge-based programs are epistemic protocol specifications in which
the set $\Phi$ is a collection of  formulas of the form $AG(x\Leftrightarrow K_i\psi)$, with exactly one such formula for each  agent $i\in \Ags$ and each template variable $x\in \Vars(\KProt_i)$.   
That is, each template variable is associated with a formula of the form $K_i\psi$, expressing some
property of agent $i$'s knowledge, and  we require that the meaning of the 
template variable be equivalent to this property.  The following example, an 
extension of an example from \cite{BrafmanLMS1997},  illustrates the motivations for knowledge-based programs
that have been advocated in the literature.

\begin{example} \label{two_robot}
Two robots, $A$	and $B$, sit on	linear track with discretized
positions $0\ldots 10$.  Initially $A$ is at position $0$ and $B$ is
at position $10$.  Their objective is to meet at a position at least
$2$, without colliding.	
Each robot is equipped with noisy position sensor, that gives at
each moment of time a natural number value in the interval $[0,\ldots,
10]$. (We consider various different sensor models below, each defined by a 
relationship between the sensor reading and the actual position.) 
The robots do not have a sensor	for detecting each other's position.
Each robot has an action $\act{Halt}$ and 
an 
action $\act{Move}$. The $\act{Halt}$ action brings the
robot to a stop at its current location, and it will not move again
after this action has been performed.  The $\act{Move}$ action moves the
robot in the direction that it is facing (right, i.e., from 0 to 10
for $A$, and left for $B$). However, the effects of this action	are unreliable:	
when performed, the robot either stays at its current position	
or moves one step in the	designated direction.

Because of the nondeterminism in the sensor readings
and the robot motion, it is a non-trivial matter to 
program the robots to achieve their goal. In particular,  the programmer
needs to reason about how the sensor readings are related
to the actual positions, in view of the assumptions about the possible
robot motions. However, there is a natural abstract 
description of the solution to the problem at the level of 
agent knowledge, which we may capture as a knowledge-based program 
as follows: $A$	
has the epistemic protocol specification
$$ 
\begin{array}[b]{rl} 
\KProt_A = & \pdo \\ 
& ~~ \neg x \rightarrow \act{Move} \\ 
& ~~[]  ~x ~\rightarrow \act{Halt} \\ 
& \pdor \\ \\
& AG(x\dimp K_A(position_A \geq 2)) 
\end{array}  
$$
and $B$ has  the epistemic protocol specification
$$ 
\begin{array}[b]{rl} 
\KProt_B = & \pdo \\ 
& ~~ y \rightarrow \act{Move} \\ 
& ~~[]~\neg   y \rightarrow \act{Halt} \\ 
& \pdor \\ ~\\
& AG(y\dimp K_B(\bigwedge_{p\in [0, \ldots 10]} position_B = p \rimp 
AG (position_A < p -1)))
\end{array}  
$$
Intuitively, the specification for $A$ says that $A$ should move to the right until it knows
that its position is at least 2. The specification for $B$ says that $B$ should move
to the left so long as its knows that, if its current position is $p$, 
then 
$A$'s position will always be to the left of the position $p-1$ that a move 
might cause $B$ to enter. If this does not hold then there could be a collision. 

One of the benefits of knowledge-based programs is that they can be 
shown to guarantee correctness properties of solutions for a problem independently of the 
way that knowledge is acquired and represented. This gives a 
desirable level of abstraction that enables a single knowledge
level description to be used to generate multiple implementations that are 
tailored to different environments. 

In the case of the above knowledge-based program, 
we note that it  guarantees several properties 
independently of the details of the sensor model. Informally, 
since $A$ halts only when it knows that its position 
is at least 2, and $K_A p \rimp p$ is a tautology of the logic of knowledge, 
its program ensures that when $A$ halts, its position will be at least 2. 
Similarly, since $B$ moves at most one position in any step, 
and moves only when it knows that moving to the position to its left will not cause a collision 
with $A$, a move by $B$ will not be the cause of a collision. 
It remains to show that $A$ does not cause a collision with  
$B$ --- this requires assumptions about $A$'s sensor. 
(Note that if A is blind it never halts, and could collide with B even if $B$ never moves, so assumptions are needed.) 
For termination, moreover, we require fairness assumptions about the way that 
$A$ and $B$ move (e.g., an action $\act{Move}$   performed infinitely often
eventually causes the position to change.).

What implementations exist for the knowledge-based program depend on 
the assumptions we make about the error in the sensor readings. 
We assume that for each agent $i$, and possible sensor value $v$,
there are propositions $sensor_i = v$, $sensor_i \geq v$, and $sensor_i \leq v$ in $\Prop_i$, with the obvious meaning. 
Suppose that we take the robots' position sensor to
be free of error, i.e. for each agent $i$, we always have $sensor_i = position_i$. 
Then agent $i$ always knows its exact position from its sensor value. 
In this case, the knowledge-based program has an implementation
with $\theta(x)$ is $sensor_A= 2$ and $\theta(y)$ is $sensor_B \geq 4$. 
In this implementation, $A$ halts at position 2 and $B$ halts 
at position 3 (assuming that they reach these positions.) 

On the other hand, suppose that the sensor readings may be erroneous, with 
a maximal error of 1, i.e., when the robot's position is $p$, the sensor value 
is in $\{p-1,p,p+1\}$. 
In this case, there exists an implementation $\theta$ in which 
$\theta(x)$ is $sensor_A = 3 \lor sensor_A = 4 \lor sensor_A = 5$, 
and
 $\theta(y)$ is $ sensor_B = 4 \lor sensor_B = 5 \lor sensor_B= 6$. 
In this implementation, $A$ moves until it gets a sensor reading in the set $\{3,4,5\}$, and then halts. 
The effect is that $A$ halts at a location in the set $\{2,3,4\}$; which one
depends on the pattern of sensor readings obtained. For 
example, the sequence $(0,0), (1,1),(2,2),(3,2),(4,3)$ of $(position, sensor)$ values 
leaves $A$ at position 4, whereas
the sequence $(0,0), (1,1),(2,3)$  leaves $A$ at position $2$. 
The effect of the choice of $\theta(y)$ is that $B$ moves to the left and halts in one of the positions $\{5,6,7\}$. 
One run in which $B$ halts at position $5$ has $(position, sensor)$ values
$(10,10), (9,9),(8,8),(7,7),(6,7),(5,4)$. 
A run in which $B$ halts at position $7$ is where these values are 
$(10,10), (9,9),(8,8),(7,6)$. Note that here the sensor reading 6 
tells $B$ that it is in the interval $[5,7]$, so it could be at $5$. It is therefore not safe to move, 
since $A$ might be at $4$.

One of the advantages of the knowledge-based programs is that their 
implementations are optimal in the way that they use the information
encoded in the agent's observations. For example, the 
program for $A$  says that 
$A$ should halt \emph{as soon} as it knows that it is in the goal region. 
In the case of the sensor with noise at most 1, the putative implementation for $A$ given 
by $\theta(x) = sensor_A \geq 4$ would also ensure that $A$ halts inside the goal region $[2,10]$, but would 
not implement the knowledge-based program because there are situations (viz. $sensor_A = 3$), 
where $A$ does not halt even though it knows that it is safe to halt. 
\end{example}

The semantics for knowledge-based programs results in implementations that are highly optimized in their use of information. 
Because knowledge for an implementation $\theta$ is computed in the system $\I(E, P\theta)$, 
agent's may reason with complete information about the implementation  they are running in determining what information follows from their
observations. This introduces a circularity that makes finding implementations  of knowledge-based programs an inherently complex problem. 
Indeed, it also has the consequence that it is possible for a knowledge-based program to have no implementations. The following
provides a simple example where this is the case. It also illustrates 
a somewhat counterintuitive aspect of knowledge-based programs, that we will argue 
is improved by our proposed ordered semantics for epistemic specifications below. 

\begin{example} \label{ex:picnic} 
Alice and Bob have arranged to meet for a picnic. They are agreed that a picnic
should have both wine and cheese, and each should bring one or the other. However, they did not
think to coordinate in advance what each is bringing, and they are now not able to communicate, 
since Alice's phone is in the shop for repairs. They do know that each reasons as follows. 
Cheese being cheaper than wine, they prefer to bring cheese, and will do so if they 
know that there is already guaranteed to be wine. Otherwise, they will bring wine. This situation 
can be captured by the knowledge-based program (for each $i\in \{A,B\}$) 
and environment depicted in Figure~\ref{fig:picnicenv}. 
\begin{figure}[t] 
\centerline{ 
$ 
\begin{array}[b]{l} 
\KProt_i = \pdo \\ 
~~ ~~\mathit{start} \land x_i \rightarrow \act{c} \\ 
~~[] ~ \mathit{start} \land \neg x_i \rightarrow \act{w} \\ 
~~[] ~ \neg \mathit{start} \rightarrow \act{p} \\ 
\pdor \\ ~\\
AG(x_i\dimp K_iAX w) 
\end{array} 
$ \hspace{2cm} \includegraphics[height=2.5cm]{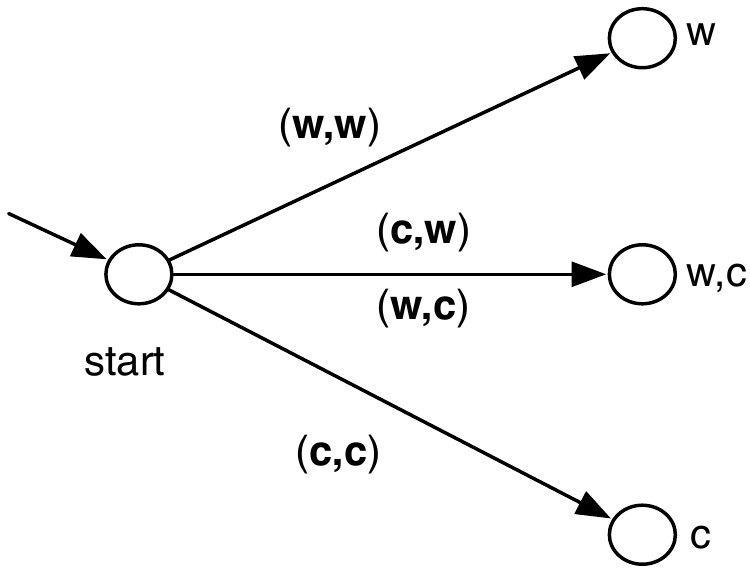}\\[-5pt]}
\caption{Knowledge-based program and environment\label{fig:picnicenv}} 
\end{figure} 
Here $\mathit{start}$ is a proposition, local to both agents, that holds before the picnic (at time 0).
We use $w,c$ as propositions that hold if there is wine (respectively, cheese) in the picnic state (at time 1). 
Actions $\act{w},\act{c}, \act{p}$ represent bringing wine, bringing cheese, and picnicking, respectively.  
For any omitted joint actions $\act{a}$ from a state $s$ in the diagram, we assume an implicit
self-loop $s \ptrans{\act{a}} s$. 
We assume that for all states $s$ and $i\in \{A, B\}$, we have $O_i(s) = s$, i.e., both agents
have complete information about the current state. 

This epistemic specification has no implementations. Note that in any implementation, each agent $i$ 
must choose either $\act{w}$ or $\act{c}$ at the initial state. For each such selection, there is a unique successor state
at time 1, so each implementation system $\I(\KProt\theta, E)$ has exactly one state at time 1. If this state satisfies 
$w$, then we have $\I(\KProt\theta, E) \models K_i(AX w)$, and this implies that both agents select action $\act{c}$ at the start state. 
But then the state at time 1 does not satisfy $w$. Conversely, if the unique state at time $1$ does not satisfy $w$, then 
$\I(\KProt\theta, E) \models \neg K_i(AX w)$, and this implies that both agents select action $\act{w}$ at the start state, which produces
a state at time 1 that satisfies $w$, also a contradiction. In either case, the assumption that we have an implementation 
results in a contradiction, so there are no implementations. \qed
\end{example}

Testing whether there exists an implementation of a knowledge-based program when the 
temporal basis of the temporal epistemic logic used is the linear time logic LTL is PSPACE 
complete \cite{FHMV1997}. However, the primary source of the hardness here 
is that model checking LTL is already a PSPACE complete problem. 

In the case of CTL as the temporal basis, where model checking can be done in PTIME, 
the problem of deciding the existence of an implementation of a given knowledge-based program in a given environment 
can be shown to be NP-complete.  NP hardness follows from Theorem 5.4 in \cite{FHMV1997}, which  states
that for \emph{atemporal} knowledge-based programs, in which the knowledge formulas $K_i \phi$ 
used do not contain temporal operators, the complexity of determining the existence of
an implementation is NP-complete. However, the construction in the proof in \cite{FHMV1997} requires both the 
environment and the knowledge-based program to vary. 
In practice, the size of the knowledge-based program is likely to be significantly smaller than the size of the 
environment, inasmuch as it is created by hand and effectively amounts to a form of specification. 
An alternate approach  is to measure complexity as a function of the size of the environment for a fixed knowledge-based program.    
Even here, it turns out, the problem of deciding the existence of an implementation 
is NP-hard for very simple knowledge-based programs. 

\begin{theorem} \label{thm:atemp}
There exists a fixed atemporal knowledge-based program $\KProt$ for a single agent, such that the problem of 
deciding, given an environment $\Env$, whether $\KProt$ has an implementation in $\Env$, is NP-hard. 
\end{theorem} 

The upper bound of NP for deciding the existence of implementations of knowledge-based programs 
is generalized by the following result for our more general notion of epistemic protocol specification. 

\begin{theorem} \label{thm:eps-upper} 
Given an environment $\Env$ and an epistemic protocol specification $\EPS$ expressed using $\CTLK$, the complexity of 
determining the existence of an implementation for $\EPS$ in $\Env$ is in NP. 
\end{theorem} 

Theorem~\ref{thm:eps-upper} assumes that the environment is presented by means of 
an explicit listing of its states and transitions. 
In practice, the inputs to the problem will be given in 
some format that makes their representation succinct, 
e.g., states will be represented as assignments to some
set of variables, and boolean formulas will be used to 
represent the environment and protocol components. 
For this alternate input format, the problem of determining the existence of an implementation of a given 
epistemic protocol specification is NEXPTIME-complete 
\cite{HM14tacas}.  

Under either an implicit or explicit representation of environments, these results
suggest that synthesis of implementations of general epistemic protocol specifications, 
and knowledge-based programs in particular, 
is unlikely to be practical. An implementation using symbolic techniques is 
presented in \cite{HM14tacas}, but it works only on small examples and scales poorly
(it requires the introduction of exponentially many fresh propositions before using
BDD techniques; the number of propositions soon reaches the limit that can be handled efficiently by 
BDD packages.) 
In the following section, we consider a restricted class of specifications that weakens
the notion of knowledge-based program in such a way
that implementations can always be found, and focus on how to efficiently derive 
implementations that approximate the implementations of corresponding
knowledge-based programs as closely as possible. 

 \section{An Ordered Semantics} \label{sec:ordsem}

Sound local proposition epistemic protocol specifications are 
a generalization of knowledge-based programs, introduced in 
\cite{EMM98}, with one of the motivations being that they 
provide a larger space of potential implementations, 
that may overcome the problem of the high complexity of
finding an implementation.  (There is the further motivation that the 
implementation of a knowledge-based program, when one exists,  itself may be intractable; e.g., 
it is shown  in \cite{Meyden96} that for perfect recall implementations of \emph{atemporal} 
knowledge-based programs, deciding whether $K_i\phi$ holds at a given 
point of the implementation may be a PSPACE-complete problem. This specific
motivation is less of concern for the observational case that we study in this paper.) 

Formally, a \emph{sound local proposition} epistemic protocol specification is one in which $\Phi$ is 
given by means of a function $\kappa$ with domain $\Vars(\KProt)$, such that for each agent $i$ and each template variable 
$x\in \Vars(\KProt_i)$, the formula $\kappa(x)$ is of the form $K_i \psi$. The corresponding
set of formulas for the epistemic protocol specification is $\Phi = \Phi_\kappa = \{ AG(x \rimp \kappa(x)) ~|~x\in 
\Vars(\KProt) \}$.

As usual for  epistemic protocol specifications, an implementation associates 
to each template variable a boolean formula local to the corresponding agent,  
such that the resulting system satisfies the specification $\Phi$.%
\footnote{By the assumption of locality of $\theta(x)$, 
validity of $AG(\theta(x)\Rightarrow K_i\psi)$ in a system  is equivalent to validity of 
$AG(\theta(x)\Rightarrow \psi)$, 
but we retain the epistemic form for emphasis and to maintain the connection to knowledge-based programs.}  Thus, whereas a knowledge-based program requires
that each knowledge formula in the program be implemented by a {\em necessary and sufficient} 
local formula, a sound local proposition specification requires only that the implementing
local formula be \emph{sufficient}. 

It is argued in \cite{EMM98} that examples of knowledge-based programs 
can typically be weakened to sound local proposition specifications 
without loss of the desired \emph{correctness} properties that hold of all implementations. 
However, implementations of knowledge-based programs may guarantee 
\emph{optimality} properties that are not guaranteed by the corresponding sound local proposition specifications. 
For example, an implementation of a knowledge-based program that states ``if $K_i \phi$ then do $a$" will 
be optimal in the sense that it ensures that the agent will do $a$ \emph{as soon as} it knows that $\phi$ holds.
By contrast, an implementation that replaces $K_i\phi$ by a sufficient condition for this formula 
may perform $a$ only much later, or even fail to do so, even if the knowledge necessary to do $a$ 
is deducible from the agent's local state. (An example of such a situation is given in \cite{BaukusMeyden}, which identifies 
a situation where a cache coherency protocol fails to act on knowledge that it has.) 

Note that  the substitution $\theta_\bot$, defined by $\theta_\bot(x) = \mathbf{false}$ for all template variables 
$x$, is \emph{always} an implementation for a sound local proposition specification   $\EPS$
in an environment $\Env$.  It is therefore trivial to decide the existence of
an implementation, and it is also trivial to produce a succinct representation of an implementation. 
Of course, an implementation of a program ``if $x$ then do $a$" that sets $x$ to be $\mathbf{false}$ will 
never perform $a$, so this trivial implementation is generally not of much interest.
What is more interesting is to find \emph{good} implementations, that approximate the 
corresponding knowledge-based program implementations as closely as possible in order to behave 
as close to optimally as possible, while remaining tractable. 

Consider the order on substitutions defined by $\theta \leq \theta'$ if 
for all variables $x$ and states $s\in S$ of the environment 
we have $\pi(s) \models \theta(x) \rimp \theta'(x)$. 
If both are implementations of $\EPS$ in $\Env$, we may find 
$\theta'$ preferable in that it provides weaker sufficient conditions (i.e., ones more often true) 
for the knowledge formulas $K_i\phi$ of interest. 
Pragmatically, if $\phi$ is a condition that an agent must know
to be true before it can safely perform a certain action,  the
more often the sufficient condition $\theta(x)$ for $K_i\phi$ holds, 
the more often will the agent perform the action in the implementation. 
It is therefore reasonable to seek implementations that maximize 
$\theta$ with respect to the order~$\leq$. 
The maximal sufficient condition for $K_i\phi$ is $K_i\phi$ itself, in the system 
$\I(\Env, P\theta)$ corresponding to an implementation $\theta$, expressed as an equivalent  local formula.%
\footnote{
The existence of such a formula follows from completeness of the set of local propositions. 
If we extend the propositions in an 
environment to include for each agent $i$ and possible observation 
$o$ of the agent, a  proposition $p_{i,o}$ that holds at a state $s$ iff
$O_i(s) =o$, then the formula $\theta(x)$ such that $\I \models AG(\theta(x) \dimp \kappa(x))$, 
where $\kappa(x) = K_i \phi$,  can be constructed as $\bigvee \{ p_{i,o}~|~o \in O_i(S), ~ \I, o \models \kappa(x)\}$, 
and has size of order the number of observations.} 

The following result makes this statement precise: 

\begin{theorem} \label{thm:epskbp}
Suppose that $\EPS$ is a sound local proposition epistemic protocol specification, 
and let $\EPS'$ be the knowledge-based program resulting from replacing each 
formula $AG(x \rimp \kappa(x))$ in $\Phi$ by the formula $AG(x \dimp \kappa(x))$. 
Then every implementation $\theta$ of $\EPS'$ is an implementation of $\EPS$. 
\end{theorem} 

However, to have $\theta(x)$ equivalent to $K_i\phi$ in $\I(\Env, P\theta)$  would mean that $\theta$ implements a knowledge-based program. 
The complexity results of the previous section indicate that this is too strong a requirement, for practical purposes, since it
is unlikely to be efficiently implementable. The compromise we explore in this paper is to require 
$\theta(x)$ to be equivalent to $K_i\phi$ not in the system $\I(\Env, P\theta)$ itself, but in 
another system that approximates $\I(\Env, P\theta)$. The basis for the correctness of 
this idea is the following lemma. 

\begin{lemma} \label{lem:subsysk} 
Suppose that $\I \subseteq \I'$, that $r$ is a run of $\I$ and
that $\phi$ is a formula in which knowledge operators and the branching operator $A$ occur 
only in positive position. Then $\I',(r,m) \models \phi$ implies 
$\I,(r,m) \models \phi$.  
\end{lemma} 
 
In particular, if, for a sound local proposition epistemic protocol specification  $\EPS$, 
the formula $\kappa(x)$ associated to a template variable $x$ is in $\CTLK^+$, then
this result applies to the formula $AG(x \rimp \kappa(x)) $ in $\Phi_\kappa$, since this
is also in $\CTLK^+$. Suppose the  system $\I'$  approximates the ultimate implementation 
$\I(\Env, P\theta)$ in the sense that $\I' \supseteq \I(\Env, P\theta)$. Let $\theta(x)$ be a 
local 
formula
 such that $\I' \models AG(\theta(x) \dimp \kappa(x))$. Then also $\I' \models AG(\theta(x) \rimp \kappa(x))$, hence, by Lemma~\ref{lem:subsysk}, 
$\theta(x)$ will also satisfy the correctness condition $\I(\Env, P\theta) \models AG(\theta(x) \rimp \kappa(x))$
necessary for $\theta$ to be an implementation of $\EPS$. 

Our approach to constructing good implementations of $\EPS$ will be to 
compute local formulas $\theta(x)$ that are equivalent to $\kappa(x)$ in 
approximations $\I'$ of the ultimate implementation being constructed. 
We take this idea one step further. Suppose that we have used this 
technique to determine the  value of $\theta(x)$ for some of the 
template variables $x$ of $\EPS$. Then we have increased our information 
about the final implementation $\theta$, so we are able to construct
a \emph{better} approximation $\I''$ to the final implementation $\I(\Env, P\theta)$, 
in the sense that $\I' \supseteq \I'' \supset \I(\Env, P\theta)$. 
Note that if  $\I' \models AG(\phi' \dimp \kappa(y))$ and 
$\I'' \models AG(\phi'' \dimp \kappa(y))$, then it follows from $\I' \supseteq \I''$
that $\I'' \models AG(\phi' \rimp \phi'')$. That is, $\phi''$ is weaker than 
$\phi'$, and hence a better approximation to the knowledge condition $\kappa(y)$ in 
the ultimate implementation $\I(\Env, P\theta)$. Thus, by proceeding
iteratively through the template variables, and  improving the approximation as we 
construct a partial implementation, we are able to obtain better approximations
to $\kappa(y)$ in $\I(\Env, P\theta)$ for later variables. 

More precisely, suppose that we have a total pre-order on the set of all 
template variables $\Vars(\KProt) = \cup_{i \in \Ags} \Vars(\KProt_i)$, 
i.e., a binary relation $\leq $ on this set that is transitive
and satisfies $x\leq y \lor y\leq x$ for all $x,y\in \Vars(\KProt)$. 
Let this be represented by the sequence of subsets
$X_1, \ldots , X_k$, 
where for $i\leq j$ and $x\in X_i$ and $y\in X_j$
we have $x< y$ if $i<j$ and $x \leq y \leq x$ if $i=j$. 
Suppose we have a sequence of interpreted systems 
$\I_0 \supseteq \ldots \supseteq \I_k$. Define a substitution $\theta$ to be 
\emph{consistent} with this sequence if 
for all $i= 1\ldots k$ and $x\in X_i$, 
we have $\I_{i-1} \models AG(\theta(x) \dimp \kappa(x))$. That is, 
consistent substitutions associate to each template variable $x$ 
a local formula that is equivalent to (not just sufficient for) $\kappa(x)$, but  
in an associated approximation system rather than in the final implementation. 

\begin{proposition} \label{prop:approx}
Suppose that $\I_k $ is isomorphic to $\I(\Env,   P\theta)$, 
and that for all $x\in \Vars(\KProt)$, the formula $\kappa(x)$ contains
knowledge operators and the branching operator $A$ only in positive position. 
Then $\theta$ implements the epistemic protocol specification 
$\langle \Ags, \Env, P, \Spec_\kappa \rangle$.
\end{proposition} 
 
We will apply this result as follows: define an \emph{approximation scheme}
to be a mapping that, given an epistemic protocol specification $\EPS= \langle \Ags, \Env, P,\Phi\rangle$ 
and a partial substitution $\theta$ for $\EPS$, yields a system 
$\I(\EPS, \theta)$, satisfying the conditions
\be
\item if $\theta \subseteq \theta'$ then $\I(\EPS, \theta) \supseteq \I(\EPS, \theta')$, and 
\item if  $\theta$ is total, then $\I(\EPS, \theta)$ is isomorphic to $\I(\Env, P \theta)$. 
\ee 

Assume now that $\EPS$ is a sound local proposition specification based on 
the mapping $\kappa$. Given the ordering $\leq$ on  
$\Vars(\KProt)$,
with the associated sequence of sets
$X_1 \ldots X_k$, we define the sequence $\theta_0, \theta_1, \ldots, \theta_k$ 
inductively by $\theta_0 = \emptyset$ (the partial substitution that is nowhere defined),
and $\theta_{j+1}$ to be the extension of $\theta_j$ obtained by
defining, for $x\in X_{j+1}$, the value of $\theta_{j+1}(x)$ to be 
the local proposition $\phi$ such that $\I(\EPS, \theta_j) \models AG( \phi \dimp \kappa(x))$. 
Plainly $\theta_0 \subseteq \theta_1 \subseteq \ldots \subseteq \theta_k$, 
so we have $\I(\EPS, \theta_0) \supseteq \I(\EPS, \theta_1) \supseteq \ldots  \supseteq \I(\EPS, \theta'_k)$. 
It follows from the properties of the approximation scheme and Proposition~\ref{prop:approx} 
that the substitution $\theta_k$ is total and is an implementation of $\EPS$. 
 
This idea leads to an extension of the idea of epistemic protocol 
specifications: we now consider specifications of the form $(\EPS, \leq)$, 
where $\EPS$ is a sound local proposition epistemic protocol specification, 
and $\leq$ is a  total pre-order on the template variables of $\EPS$. 
Given an approximation scheme, the construction of the previous paragraph
yields a unique implementation of $\EPS$. Intuitively, by specifying 
an order $\leq$, the programmer fixes the order in which implementations
are synthesized for the template variables, and the approach guarantees that
variables later in the order are synthesized using information about 
the values of variables earlier in the order.

 \section{A spectrum of  approximations} \label{sec:approx} 
 
It remains to determine which approximation scheme to use in the approach to constructing
implementations described in the previous section. In this 
section, we consider a number of possibilities for the choice of approximation scheme. 
A number of criteria may be applied to the choice of approximation scheme. 
For example, since the programmer must select the order in which variables are synthesized, the 
approximation scheme should be simple enough to be comprehensible to the 
programmer, so that they may understand the consequences of their ordering 
decisions. 

On the other hand, since synthesis is to be automated, we
would like the computation of the values $\theta(x)$ to be efficient. 
This amounts to efficiency of the model checking problem 
$\I(\EPS, \theta')\models \kappa(x)$ for partial substitutions $\theta'$ and 
formulas $\kappa(x) \in \CTLK^+$. 
To analyze this complexity,  we work below with a complexity measure that assumes explicit state 
representations of environments, but we look for cases where the 
model checking problem in the approximation systems is solvable 
in PTIME. We assume that the protocol template $\KProt$  and 
the formulas $\Phi$ in the epistemic protocol specification are fixed, and measure complexity as a function of the 
size of the environment $\Env$. This is because in practice, the size of the environment is 
likely to be the dominant factor in complexity.

One immediately obvious choice for the approximation scheme 
is to take the system  $\I(\EPS, \theta)$, for a partial substitution 
$\theta$, to be the union of all the systems $ \I(E,P\theta')$,
over all total substitutions $\theta'$ that extend
the partial substitution $\theta$. 
This turns out not to be a good choice (it is the intractable case $\I_{ii,ir,sc}$ below), so we consider a number of relaxations of this
definition. The following abstract view of the situation provides a convenient
format that unifies the definition of these relaxations. 

\newcommand{\strat}{\sigma} 
\newcommand{\seq}{\rho} 

Given an environment $E$ with states $S$, define a \emph{strategy} for $E$ 
to be a function $\strat: S^+ \rightarrow \powerset{S}\setminus{\emptyset}$
mapping each nonempty sequence of states to a set of possible successors. 
We require that for each $t\in \strat(s_0\ldots s_k)$ we have $s_k \ptrans{\mathbf{a}} t$ for   
some joint action $\mathbf{a}$. Given a set $\Sigma$ of strategies, 
we can construct an interpreted system consisting of all 
runs consistent with some strategy in $\Sigma$. 
We encode the  strategy into the run.
We use the extended set of global states $S \times \Sigma$.  
We take 
$\R_\Sigma$
 to be the set of all $r: \nat \rightarrow S \times \Sigma$ 
such that  there exists a strategy $\strat$ such that for
all $n\in \nat$ we have $r(n) = (s_n, \strat)$, for some $s_n \in S$,  
and, we have $s_{n+1} \in \strat(s_0 s_1 \ldots s_n) $  for all $n \in \nat$. 
Intuitively, this is the  set of all infinite runs, each using some fixed strategy in $\Sigma$, 
with the strategy encoded into the state.  We define $\I(E, \Sigma) = (\R_\Sigma, \sim, \pi') $ 
where  $\sim = \{\sim_i\}_{i\in \Ags}$ is the relation on points of $\R_\Sigma$ defined by $(r,m) \sim_i (r',m')$ 
if, with $r(m) = (s, \strat)$ and $r'(m') = (s' ,\strat')$, we have $O_i(s) = O_i(s')$. 
The interpretation $\pi'$  on $S \times \Sigma$ is
defined so that $\pi'(s,\strat) = \pi(s)$, where $s \in S$, $\strat \in \Sigma$ and 
$\pi$ is the interpretation from $\Env$.

\newcommand{\pinf}{\mathit{pi}} 
\newcommand{\ii}{\mathit{ii}} 

\newcommand{\pr}{\mathit{pr}} 
\newcommand{\ir}{\mathit{ir}} 

A \emph{memory definition} is a collection of functions $\mu = \{\mu_i\}_{i\in \Ags}$
with each $\mu_i$ having domain $S^+$.  
In particular, we work with the following memory definitions derived using the observation functions in the 
environment $E$: 
\begin{itemize} 
\item The {\em perfect information, perfect recall} definition $\mu^{\pinf,\pr} = \{\mu^{\pinf,\pr}_i\}_{i\in \Ags}$ where \\
$\mu_i^{\pinf,\pr}(s_0\ldots s_k) =  s_0\ldots s_k$

\item The {\em perfect information, imperfect recall} definition $\mu^{\pinf,\ir} = \{\mu^{\pinf,\ir}_i\}_{i\in \Ags}$ where \\
$\mu_i^{\pinf,\ir}(s_0\ldots s_k) =  s_k$

\item The {\em imperfect information, perfect recall} definition $\mu^{\ii,\pr} = \{\mu^{\ii,\pr}_i\}_{i\in \Ags}$ where \\
$\mu_i^{\ii,\pr}(s_0\ldots s_k) =  O_i(s_0)\ldots O_i(s_k)$

\item The {\em imperfect information, imperfect recall} definition $\mu^{\ii,\ir} = \{\mu^{\ii,\ir}_i\}_{i\in \Ags}$ where \\
$\mu_i^{\ii,\ir}(s_0\ldots s_k) =  O_i(s_k)$
\end{itemize} 
A strategy \emph{depends} on memory definition $\mu$ 
if there exist functions $F_i: \mathit{range}(\mu_i)  \rightarrow \powerset{\Acts_i}$ for $i \in \Ags$
such  that for all sequences $\rho= s_0 \ldots s_k$, 
we have $t\in \strat(s_0 \ldots s_k)$ iff 
$s \ptrans{\mathbf{a}} t$ for some  joint action $\mathbf{a}$  
such that for all 
$i\in \Ags$, we have $\mathbf{a}_i \in  F_i(\mu_i(s_0 \ldots s_k))$.

Let $\KProt$ be a joint protocol template and let $\theta$ be a partial substitution for $\KProt$. 
A strategy $\strat$ is \emph{substitution consistent} 
with respect to  $\KProt,\theta$ and a memory definition $\mu$ if 
$\strat$ depends on $\mu$ and for all sequences $s_0 \ldots s_k$ 
there exists a substitution $\theta'\supseteq \theta$ mapping all the template variables of $\KProt$ undefined by 
$\theta$ to truth values, such that 
\begin{equation}\label{eq:subcons} 
\strat(s_0 \ldots s_k) = \{t ~|~\text{there exists } \act{a} \in \enabled(P \theta', s_k), ~ s_k \ptrans{\act{a}} t \}
\end{equation} 
Note that since the choice of $\theta'$ is allowed to depend on $s_0 \ldots s_k$, 
this does not imply that the set of possible successors states $\strat(s_0 \ldots s_k)$  depends
only on the final state $s_k$;  the reference to $s_k$ in the right hand side of 
equation~\ref{eq:subcons} 
is  included just to allow the enabled actions to be determined in a way
consistent with the substitution $\theta$, which already associates
some of the variables with predicates on the state $s_k$.

\begin{example} \label{ex:top-nsc}
Consider the maximally nondeterministic, or  \emph{top},  strategy $\strat_\top$, 
defined by $\strat_\top(s_0\ldots s_k) = \{ t ~|~ \text{there exists $\act{a}\in \Acts$} , s_k \ptrans{\act{a}} t\}$ for all $s_0\ldots s_k$. 
Intuitively, this strategy allows any action to be taken at any time. 
It is easily seen that $\strat_\top$ depends on every memory definition $\mu$. 
However, it is not in general substitution consistent, since there are protocol templates
for which the set of enabled actions (and hence the transitions) 
depend on the substitution. 

Consider the protocol template $\KProt = \pdo~ x \rightarrow a~ []~ \neg x \rightarrow b ~\pdor$
for a single agent,  in an environment with states $S =  \{s_0,s_1,s_2\}$ 
and transitions $s_0 \ptrans{a} s_1$, $s_0 \ptrans{b} s_2$, $s_1 \ptrans{a,b} s_1$ and $s_2 \ptrans{a,b} s_2$. 
Let $\theta$ be the empty substitution. For all substitutions $\theta'$, $\enabled(\KProt\theta', s_0)$ is either $\{a\}$ or $\{b\}$, 
so for the sequence $s_0$, the right hand side of equation~(\ref{eq:subcons})  is 
equal to either $\{s_1\}$ or $\{s_2\}$. For the strategy $\sigma_\top$, we have 
$\sigma_{\top}(s_0) = \{s_1,s_2\}$. Hence this strategy is not substitution consistent in this environment. \qed
\end{example}

\newcommand{\subc}{\mathit{sc}}
\newcommand{\nsubc}{\mathit{nsc}}

We now obtain eight sets of strategies 
by choosing an information mode $a \in \{\pinf,\ii\}$, 
a recall mode $b\in \{\pr,\ir\}$ and a selection $c\in \{\subc,\nsubc\}$ 
to reflect a choice with respect to the requirement of substitution consistency.  
Formally, given a joint protocol template $\KProt$, a partial substitution $\theta$ for $\KProt$, 
and an environment $\Env$, we define $\Sigma^{a,b,c}(\KProt,\theta, \Env)$ 
to be the set of all strategies in $\Env$ that depend on $\mu^{a,b}$, 
and that are substitution consistent with respect to $P, \theta$ and $\mu^{a,b}$ 
in the case $c = \subc$.  

Corresponding to these eight sets of strategies, we obtain 
eight 
approximation schemes. Let $\EPS$ be an epistemic protocol
specification with joint protocol template $\KProt$, and environment $\Env$. 
Given a partial substitution $\theta$ for $\KProt$, and a triple $a,b,c$, 
we define the system $\I_{a,b,c}(\EPS, \theta)$ 
to be $\I(\Sigma^{a,b,c}(P, \theta,E),E)$.

\begin{proposition}
For each   information mode $a \in \{\pinf,\ii\}$, 
a recall mode $b\in \{\pr,\ir\}$ and selection $c\in \{\subc,\nsubc\}$, 
the mapping $\I_{a,b,c}$ 
is an approximation scheme. 
\end{proposition} 

Additionally we have the approximation scheme 
$\I^{\top}(\EPS, \theta)$ 
defined to be $\I(\{\strat^\top_{\Env, P\theta}\},\Env)$, 
based on the top strategy  in $\Env$ relative to 
the protocol template $\KProt\theta$,  
which is defined by taking
$\strat^\top_{\Env, \KProt\theta} (s_0 \ldots s_k)$ 
to be the set of all states $t\in S$ such that there exists a joint action  $a\in \Acts$
such that for all $ i \in \Ags$,  the protocol template $\KProt_i\theta$ 
contains a clause $\phi \theta \rightarrow a_i$ with $\phi\theta$ 
satisfiable relative to $\pi(s_k)$. 
(We note that here $\pi(s_k)$ provides the values of propositions $\Prop$
and we are asking for satisfiability for some assignment to the variables 
on which $\theta$ is undefined. Because we are interested in the 
case where $\KProt$, and hence $\phi$, is fixed, 
this satisfiability test can be performed in PTIME as the environment
varies.)

For reasons indicated in Example~\ref{ex:top-nsc}, the 
strategy $\strat^\top_{\Env, P\theta}$ is not substitution-consistent. 
However, it is easily seen to depend only on the values $O_i(s_k)$, so 
we have $\strat^\top_{\Env, P\theta} \in \Sigma^{ii,ir,nsc}$.  

\begin{figure} 
\centerline{\includegraphics[height=7cm]{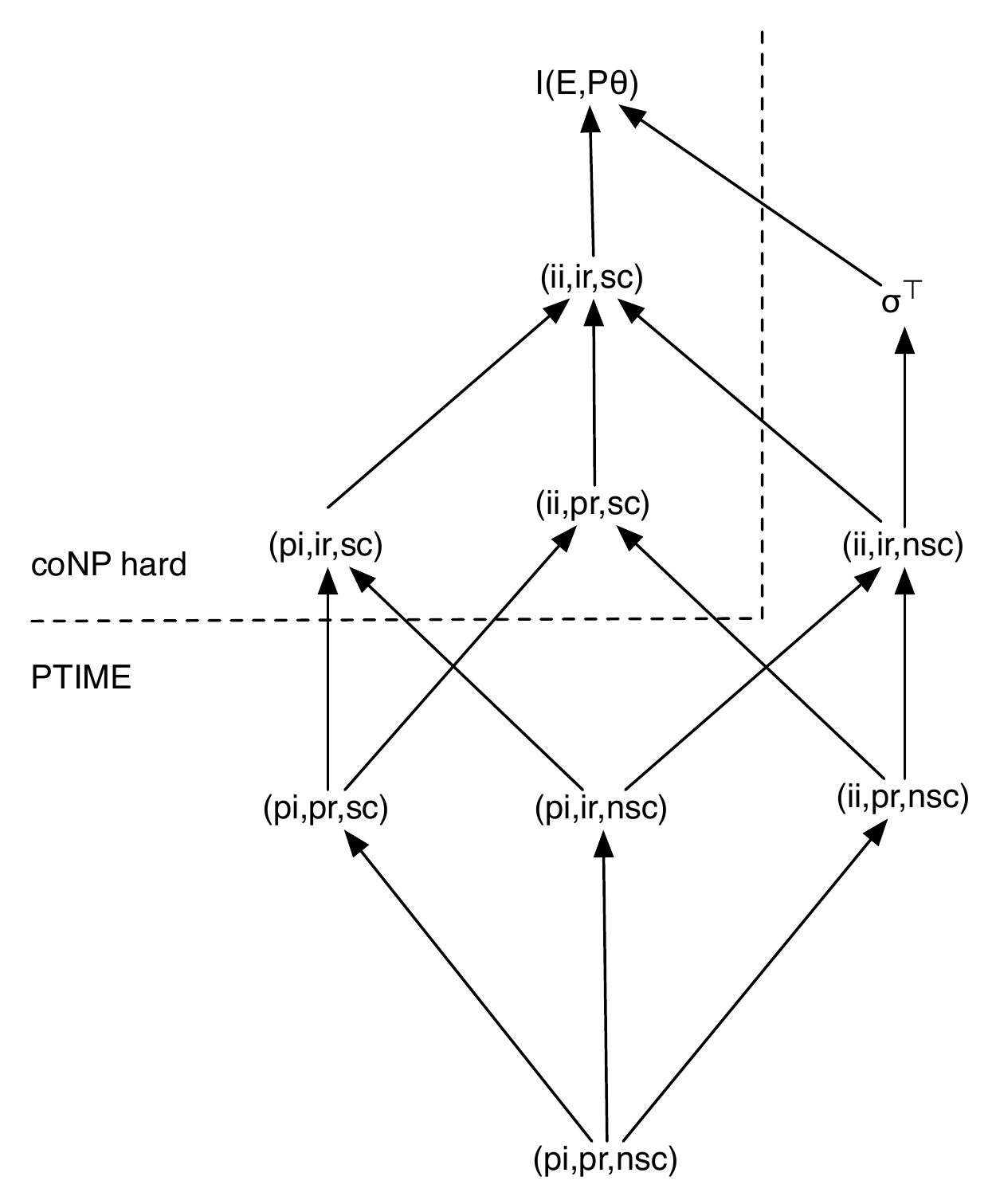}}
\caption{Lattice structure of the approximations\label{fig:lattice}} 
\end{figure}

Figure~\ref{fig:lattice} shows the lattice structure of the approximation
schemes, with an edge from a scheme $\I$ to a scheme $\I'$ 
meaning that $\I'$ is a closer approximation to the final system 
$\I(E, \KProt\theta)$ synthesized, informally in the sense that $\I$ has more runs and more branches from 
any point  than does $\I'$. (Generally, the relation is one of simple containment of the sets of runs, but in the 
case of edges involving $\I(E, \KProt\theta)$  and $\strat^\top$, we need a notion of 
simulation to make this precise.)  

Besides yielding an approach to the construction of implementations of 
epistemic protocol specifications, we note that our approach 
also overcomes the counterintuitive aspect of knowledge-based programs
illustrated in Example~\ref{ex:picnic}. 

\begin{example} 
Suppose that we replace the specification formulas $AG(x_i\dimp K_iAX w)$
in Example~\ref{ex:picnic} by the weaker form $AG(x_i\rimp K_iAX w)$, and 
impose the ordering $x_A< x_B$ on the template variables. 
We compute the implementation obtained when we use $\I^{\top}$ as the approximation scheme. 
We take $\theta_0$ to be the empty substitution. 
$\I(\{\strat^\top_{\Env, P\theta_0}\},\Env)$ has all possible behaviours of the original environment, 
so at the start state, we have $\neg K_A (AXw)$. It follows that substitution $\theta_1$, which 
has domain $\{x_A\}$ assigns to $x_A$ a local proposition that evaluates to $\mathbf{false}$ at the initial state. 
Hence, $\KProt_A\theta_1$ selects action $\act{w}$ at the initial state. The effect of this is to 
delete the bottom transition from the state transition diagram for the environment in Figure~\ref{fig:picnicenv}.  It follows that 
in  $\I(\{\strat^\top_{\Env, P\theta_1}\},\Env)$, we have $K_B(AX w)$ at the initial state, so $\theta_2(x_B)$
evaluates to $\mathbf{true}$ at the initial state. This means that the final implementation $\KProt \theta_2$ is 
the protocol in which Alice brings wine and Bob brings cheese, leading to a successful picnic, 
by contrast with the knowledge-based program, which does not yield any solutions to their planning problem. 
(We remark that both Alice and Bob could compute
this implementation independently, once given the ordering on the variables. They do not need
to communicate during the computation of the implementation.)
\qed
\end{example}  

We noted above in Theorem~\ref{thm:epskbp} that a sound local proposition specification
obtained from a  knowledge-based program includes, amongst 
its implementations, all the implementations of the knowledge-based program. 
The knowledge-based program, in effect, imposes additional optimality 
constraints on these implementations. Our ordered semantics aims
to approximate these optimal implementations. It is therefore of interest to determine whether the
ordered semantics for sound local proposition specifications can sometimes find such 
optimal implementations. Although it is not true in general, there are situations where 
the implementations obtained are indeed optimal. The following 
provides an example. 

\begin{example} 
Consider the sound local proposition  specification obtained from the knowledge-based program of  
Example~\ref{two_robot} by replacing the $\dimp$ operators in the formulas by $\rimp$. 
That is, we take $\Phi$ to contain the formulas
$$AG(x\rimp K_A(position_A \geq 2)) $$
and 
$$AG(y\rimp K_B(\bigwedge_{p\in [0, \ldots 10]} position_B = p \rimp AG (position_A < p -1))) $$ 
We consider the setting where sensors readings are within 1 of the actual position. 
Suppose that we use $\I^{\top}$ as the approximation scheme, and order the template variables 
using $x<y$, i.e., we synthesize a solution for $A$ before synthesizing a solution for 
$B$ (knowing what $A$ is doing.) Then, for $A$, 
we construct $\theta(x)$ as the local proposition for $A$
that satisfies 
$$AG(x\dimp K_A(position_A \geq 2)) $$
in a system where both $A$ and $B$ may choose either action 
$\act{Move}$ or $\act{Halt}$ at any time. 
We obtain the substitution $\theta_1$ where $\theta_1(x)$ is $sensor_A\geq 3$, which ensures
that always $position_A \leq 4$, and in which $A$ may halt at a position in the set $\{2,3,4\}$. 
In the next step, we synthesize $\theta(y)$ 
as the local proposition such that 
$$AG(y\dimp K_B(\bigwedge_{p\in [0, \ldots 10]} position_B = p \rimp AG (position_A < p -1))) $$ 
in the system where $A$ runs $\KProt_A\theta_1$, and where $B$ may choose either action 
$\act{Move}$ or $\act{Halt}$ at any time. 
In this system, $B$ knows that $A$'s position is always at most $4$, 
so it is safe for $B$ to move if $position_B \geq  6$. Agent $B$ knows
that its position is at least 6 when it gets a sensor reading at least 7. 
Hence, we obtain the substitution $\theta_2$  where $\theta_2(y)$ is $sensor_B\geq 7$
and $\theta_2(x)$ is $sensor_A\geq 3$. It can be verified that this  
substitution is in fact an implementation of the original knowledge-based  program. 
\end{example}

\section{Complexity of model checking in the approximations} \label{sec:complex} 

To construct an implementation based on the extended epistemic 
protocol specification $(\EPS, \leq)$ using an approximation scheme 
$\I(\EPS, \theta)$, we need to perform model checking of formulas in $\CTLK^+$ 
in the  systems produced by the approximation scheme. 
We now consider the complexity of this problem for the approximation schemes introduced in the 
previous sections. We focus on the complexity of this problem with the 
protocol template fixed as we vary the size of the environment, for reasons explained above. 

 \newcommand{\glob}{\mathit{global}}  
  \newcommand{\runencoded}{\mathit{run-enc}}

We say that the \emph{environment-complexity} of an approximation scheme 
 $\I(\EPS, \theta)$ is the maximal complexity of 
 the problem of deciding  $\I(\EPS, \theta), o \models \kappa(x)$ 
with    
all components fixed  and only the environment $\Env$ in $\EPS$
 varying. 
More precisely, write 
$\EPS^- = \langle \Ags, \KProt, \kappa \rangle$
for a tuple consisting of a set $\Ags$ of agents, a collection  $\KProt = \{\KProt_i\}_{i\in \Ags}$ of
protocol templates for these agents, and a mapping $\kappa$ associating, for each agent $i$,  
a formula $\kappa(x) = K_i\phi$ of $\CTLK^+$ to each template variable $x$ in $\KProt_i$. 
Given an environment $\Env$, write 
$\EPS^-(\Env)$ for the   epistemic protocol specification $ \langle \Ags, \Env , \{\KProt_i\}_{i\in \Ags}, \Spec_\kappa\rangle$
obtained from these components. Say that $\Env$ \emph{fits} a tuple 
$(\EPS^-,\theta, o,x)$ consisting of 
$\EPS^-$ as above, a substitution $\theta$ assigning a boolean formula to a subset of the template variables in $\KProt$, 
 an observation $o$ and a variable $x$,  if $\Env$ contains all actions used in $\KProt$, 
 $o$ is an observation in $\Env$ of the agent $i$ such that $\KProt_i$ contains $x$, and 
for each $x$ such that $\theta(x)$ is defined, the formula $\theta(x)$ is local in $E$ to the 
agent $i$ such that $\KProt_i$ contains $x$. 
Given $\EPS^-= \langle \Ags, \KProt, \kappa \rangle$ and $\theta$, 
$o$ and $x$,  
define $EC_{(\EPS^-,\theta,o,x)}$ to be
the set 
$$ \{\Env ~|~ \Env \text{ fits }(\EPS^-,\theta,o,x)  \text{ and } \I(\EPS^-(\Env),\theta), o \models \kappa (x) \} ~.$$ 
Then the environment-complexity of an approximation scheme 
 $\I(\EPS, \theta)$ is the maximal complexity of 
 the problem of deciding  the sets $EC_{(\EPS^-,\theta,o,x)}$
 over all choices of $\EPS^-$, $\theta$, $o$ and~$x$ . 
 
 We note that even though we have allowed perfect recall and/or perfect information 
 in the strategy spaces used by the approximation, when we model check in the 
 system generated by the approximation, knowledge operators are handled
 using the usual observational (imperfect recall, imperfect information) semantics. 
 The stronger capabilities of the strategies are used to  increase the size of the strategy space in order to 
weaken the approximation. (Model checking with respect to perfect recall, in particular, 
would \emph{increase} the complexity of the model checking problem, whereas we are seeking to 
decrease its complexity.) 
 
It turns out that several of the approximation schemes, that are closest 
to the final system synthesized (which would give the knowledge-based program semantics), 
share with the knowledge-based program semantics
the disadvantage of being intractable.  These are given in the following result. 

\begin{theorem} 
The approximation schemes  $\I_{ii,ir,sc}$, $\I_{ii,pr,sc}$, and $\I_{pi,ir,sc}$ have 
coNP-hard environment complexity, even for a single agent. 
\end{theorem} 

Each of these intractable cases uses substitution consistent strategies and 
uses either imperfect recall or imperfect information. The proofs vary, but
one of the key reasons for complexity in the imperfect recall cases is that the 
strategy must behave the same way each time it reaches a state. 
Intuitively, this  means that we can encode existential choices from an NP hard problem 
using the behaviour of a strategy at a state in this case. In the case of $\I_{ii,pr,sc}$, 
we use obligations on multiple branches indistinguishable to the agent to
force consistency of independent  guesses representing the same existential choice.
All the remaining approximation schemes, it turns out, are tractable: 

\begin{theorem} 
The approximation schemes  $\I^\top$, $\I_{ii,ir,nsc}$, $\I_{pi,pr,sc}$, $\I_{pi,ir,nsc}$, 
$\I_{ii,pr,nsc}$ and $\I_{pi,pr,nsc}$ have environment complexity in PTIME. 
\end{theorem} 

The reasons are varied, but there are close connections to some known results. 
The scheme $\I^\top$ effectively builds a new finite state environment from the 
environment and protocol by  allowing some transitions that would normally be 
disabled by the protocol, so its model checking problem reduces to an instance of 
CLTK model checking, which is in PTIME by a mild extension of the usual
CTL model checking approach. It turns out, moreover, by simulation arguments, that 
for model checking $\CTLK^+$ formulas, the approximations $\I_{ii,ir,nsc}$ and 
$\I_{ii,pr,nsc}$ are equivalent to $\I^\top$, i.e., satisfy the same formulas at the same states, so the 
algorithm for $\I^\top$ also resolves these cases.  

The cases $\I_{pi,pr,sc}$ and $\I_{pi,pr,nsc}$ are very close to 
the problem of \emph{module checking} of universal $\CTL$ formulas, which is 
known to be in PTIME \cite{KupfermanVW01}. The proof technique here involves
an emptiness check on a tree automaton representing the space of perfect information, 
perfect recall strategies (either substitution consistent or not required to be so), 
intersected with an automaton representing the complement of the formula. 
The cases  $\I_{pi,pr,nsc}$ and $\I_{pi,ir,sc}$ can moreover be shown to be 
equivalent by means of simulation techniques, so the latter also falls into PTIME. 

The demarcation between the PTIME and co-NP hard cases is
depicted in Figure~\ref{fig:lattice}. This shows there are two best candidates for use 
as the approximation scheme underlying our synthesis approach. 
We desire an approximation scheme that is as close as possible to the knowledge-based program semantics, 
while remaining tractable. The diagram shows two orthogonal approximation schemes that are 
maximal amongst the PTIME cases, namely $\I^\top$ and $\I^{pi,pr,sc}$. The former generates a bushy approximation 
in that it relaxes substitution consistency. The latter remains close to the original protocol by using 
substitution consistent strategies, but at the cost of allowing perfect information, perfect recall strategies. 
It is not immediately clear what the impact of these differences will be with respect to the quality of the
implementations synthesized using these schemes, and we leave this as a question for future work.

\section{Related Work} 
\label{sec:related} 

Relatively little work has been done on automated synthesis of 
implementations of knowledge-based programs or of 
sound local proposition specifications, particularly with 
respect to the observational semantics we have studied in this paper.
In addition to the works already cited above,  some papers 
\cite{Meyden96fst,Meyden96pricai,meydenvardi,MeydenWilke05,BozianuDF14} 
have studied the complexity of synthesis with respect to specifications
in temporal epistemic logic using the synchronous perfect recall semantics. 
A symbolic implementation for  knowledge-based programs that run only a finitely bounded number of 
steps under a clock or perfect recall  semantics for knowledge is developed in \cite{HMtark13}.

There also exists a line of work that is applying knowledge based approaches and model checking techniques 
to problems in discrete event control, e.g.,  \cite{BensalemPS10,GrafPQ12,KatzPS11}. 
In general, the focus of these works is more specific than ours (e.g., in restricting to synthesis for safety properties, 
rather than our quite general temporal epistemic specifications) but there is a similar use of monotonicity. 
It would be interesting to apply our techniques in this area and conduct a  comparison of the results.

\section{Conclusion}
\label{sec:concl} 

In this paper we have proposed an ordered semantics for sound local proposition epistemic 
protocol specifications, and analyzed the complexity of a model checking problem required 
to implement the approach, for a number of approximation schemes. 
This leads to the identification of two optimal approximation schemes,  $\I^\top$ and $\I^{pi,pr,sc}$ with respect to 
which the model checking problem has PTIME complexity in an explicit state
representation.  

A number of further steps are required to obtain a practical framework for synthesis. 
Ultimately, we would like to be able to implement synthesis using symbolic techniques, 
so that it can also be practicably carried out for specifications in
which the environment is given implicitly using program-like representations, 
rather than by means of an explicit enumeration of states. The complexity 
analysis in the present paper develops an  initial understanding of the 
nature of the model checking problems that may be helpful in developing 
symbolic implementations.  In the case of the approximation scheme $\I^\top$, in fact, 
the associated model checking problem amounts essentially to 
$\CTLK$ model checking in a transformed model, for which 
symbolic model checking techniques are well understood. 
In work in progress, we have developed an implementation of this
case, and we will report on our experimental findings elsewhere. 

In the case of the approximation $\I^{pi,pr,sc}$, the model checking problem is more akin to module checking, for which 
symbolic techniques are less well studied. This case represents an interesting
question for future research, as does the question of how the implementations
obtained in practice from these tractable approximations differ. 

Our examples in this paper give some initial data points that suggest both that
the ordered approach is able to construct natural implementations for 
the sound local proposition weakenings of knowledge-based programs that
lack implementations, as well as implementations of such weakenings that are 
in fact implementations of the original knowledge-based program. More case studies
are required to understand how general these phenomena are in practice. 
It would be interesting to find sufficient conditions under which the ordered approach 
is guaranteed to generate knowledge-based program implementations.

\bibliographystyle{eptcs}
\bibliography{syn}

\end{document}